\documentclass[reprint, amsmath,amssymb,aps,]{revtex4-2}

\usepackage{graphicx}
\usepackage{dcolumn}
\usepackage{bm}
\usepackage{subcaption}

\begin{document}

\preprint{APS/123-QED}

\title{Activity-Driven Dewetting and Rupture in Thin Liquid Films}
\author{Preethi M}
\author{Daniya Davis}%
\author{Bhaskar Sen Gupta}
  \email{bhaskar.sengupta@vit.ac.in}
\affiliation{%
 Department of Physics, Vellore Institute of Technology, Vellore, India
}%

\date{\today}
\begin{abstract}
	
Thin-film dewetting is classically governed by an adhesion-mediated spinodal instability in which curvature-driven diffusion controls post-rupture coarsening. We show that internal activity fundamentally restructures this instability. Using a minimal microscopic model of an active liquid film on a solid substrate, we identify a competition between active stresses and film-substrate adhesion that produces two independently regulated dynamical length scales: vertical liquid accumulation and lateral rupture propagation. While passive films exhibit universal diffusion-limited growth, $\ell_z(t)\sim t^{1/3}$, activity converts transport from curvature-controlled diffusion to persistence-driven motion, yielding a continuous increase of the coarsening exponent from $\approx 0.33$ to $\approx 0.6$. The growth law analysis shows that persistent self-propulsion introduces an advective flux that competes with curvature-induced chemical potential gradients, enhancing growth when the persistence length becomes comparable to the evolving domain size. Simultaneously, the rupture front transitions from dissipative spreading to strongly accelerated propagation approaching ballistic scaling. This decoupling shows that activity does not simply renormalize effective surface forces but generates a distinct nonequilibrium interfacial instability governed by the balance between persistence length and adhesion. The results provide a minimal physical mechanism linking classical thin-film dewetting to dewetting-like rupture observed in active and biological materials.	
\end{abstract}

\maketitle


\nocite{*}

\textit{Introduction:}-
Thin liquid films on solid substrates rupture through interfacial instabilities that are classically understood within equilibrium thin-film hydrodynamics~\cite{Xie,Thiele,Khanna,Sharma,Hongliu,Lessel,Gonzalez,Vrij, Redon1994, Reiter2000, Khanna2000}. In sufficiently thin films, long-range intermolecular forces generate a destabilizing disjoining pressure, leading to spinodal dewetting in which thickness fluctuations amplify, select a characteristic wavelength, and subsequently coarsen through curvature-driven transport. In this framework, both the onset of rupture and the post-instability growth laws are ultimately governed by thermodynamic free-energy minimization.

Active fluids, however, continuously inject energy at the microscopic scale and generate persistent stresses that have no equilibrium analogue. While bulk phase separation and collective motion in active matter are now well established~\cite{vicsek1995,czirok2000,paramesh,preethi2,ParameshwarA}, far less is understood about how activity modifies interfacial stability. In particular, it remains unclear whether dewetting in active films is merely a renormalized version of classical spinodal rupture, or whether internal activity introduces a qualitatively distinct instability mechanism.

This question is especially relevant in light of experiments on biological active materials, such as spreading cell monolayers, which exhibit hole nucleation, symmetry breaking, and dewetting-like retraction dynamics~\cite{Beaune,Douezan}. Despite their resemblance to thin-film rupture, a minimal physical framework linking such driven systems to classical dewetting theory is lacking.

Here we show that internal activity fundamentally restructures thin-film dewetting through competition with film-substrate adhesion. Using a minimal microscopic model, we demonstrate that activity does not simply accelerate rupture but reorganizes the governing transport mechanism. While passive films exhibit diffusion-limited coarsening, active stresses progressively convert curvature-controlled growth into persistence-driven transport, producing a continuous increase of the coarsening exponent and enabling strongly accelerated rupture-front propagation under weak adhesion. The vertical accumulation of liquid and the lateral spreading of rupture respond differently yet coherently to this competition, revealing a nonequilibrium interfacial instability controlled by the balance between persistence length and adhesion strength.

These results establish a minimal physical framework for active dewetting and identify activity-adhesion competition as a general mechanism by which nonequilibrium stresses reshape classical interfacial instabilities.\\

\textit{Model:}-
We study dewetting of a thin active liquid film on a solid substrate using a minimal particle-based model that incorporates three physical ingredients: interparticle cohesion generating an effective surface tension, film-substrate attraction controlling wettability, and persistent self-propulsion representing internal activity. The system consists of a thin film with periodic boundary conditions in the lateral directions and a reflecting boundary normal to the substrate.

Particles interact via a modified Lennard-Jones potential \cite{Bhattacharya,preethi,Puri-book}
\begin{equation}
	U_{\alpha\beta}(r)=4\left[\epsilon_r\left(\frac{\sigma_{\alpha\beta}}{r}\right)^{12}-\epsilon_{WA}\left(\frac{\sigma_{\alpha\beta}}{r}\right)^6\right],
\end{equation}
with cutoff $r_c=2.5\sigma$ for fluid-fluid interactions, while a larger cutoff of $r_c = 5\sigma$ is employed for substrate-fluid interactions to account for the longer-range influence of the substrate. $\epsilon_{WA}$ controls the strength of film-substrate adhesion and therefore the effective wettability. The dynamics obey overdamped Langevin equations \cite{allen1987, frenkel2002},
\begin{equation}
	\dot{\mathbf v}_i=-\frac{\nabla V_i}{m_i}-\zeta \mathbf v_i+\sqrt{\frac{2\zeta k_BT}{m_i}}\,\boldsymbol{\xi}_i(t)+\mathbf F_i^S ,
\end{equation}
so that momentum is not conserved and transport arises from diffusive and active motion rather than hydrodynamic flow.

Activity is introduced as persistent self-propulsion: each particle experiences a force
\begin{equation}
	\mathbf F_i^S=f_A\,\hat{v}_{r_c},
\end{equation}
where $\hat{v}_{r_c}$ aligns with the local direction of motion of neighboring particles. This rule generates a finite persistence time and run length without altering the system temperature, allowing activity to act solely as a source of nonequilibrium stress.

Starting from a uniform film, we follow rupture formation and subsequent coarsening. The characteristic domain size $\ell(t)$ is extracted from the first zero crossing of the equal-time spatial correlation function \cite{Daniya-Gravity,Bhattacharya1,Daniya-shear,Bhattacharya2, Davis1} of a coarse-grained density order parameter $\psi$
\begin{equation}
	C(r,t)=\langle\psi(0,t)\psi(\vec{r},t)\rangle - \langle\psi(0,t)\rangle \langle\psi(\vec{r},t)\rangle,
\end{equation}
while wettability is quantified through the apparent contact angle of dewetted droplets. Additional simulation details are provided in the Supplemental Material.\\


\begin{figure}
	\centering
	\includegraphics[width=0.49\linewidth]{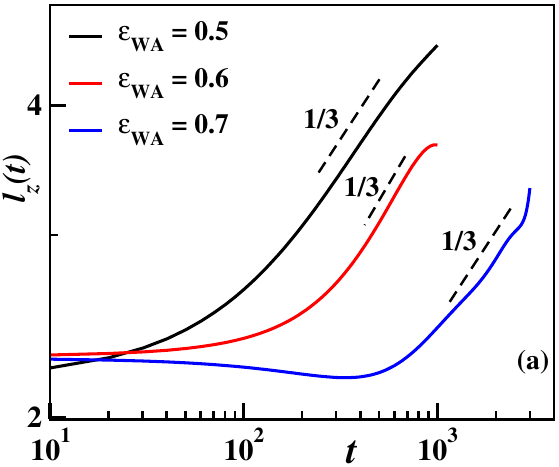}
	\includegraphics[width=0.49\linewidth]{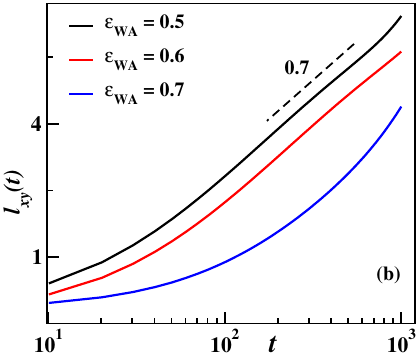}
	\caption{Growth dynamics in passive films. (a) Vertical domain size $\ell_z(t)$ showing diffusion-limited growth $\sim t^{1/3}$. (b) Lateral rupture extent $\ell_{xy}(t)$. Both observables originate from spinodal rupture, with diffusion-limited vertical growth and friction-limited contact line motion.}
	\label{fig:passive-liquid}
\end{figure}

\textit{Passive films.}-
We first establish the reference dynamics of dewetting in the absence of activity. Following rupture through the classical spinodal instability, the film evolves through two measurable length scales (Fig.~\ref{fig:passive-liquid}): the vertical liquid accumulation $\ell_z(t)$, characterizing mass redistribution normal to the substrate, and the lateral rupture extent $\ell_{xy}(t)$, describing propagation of the dry region along the surface.

The post-rupture growth of the liquid domains follows $\ell_z(t)\sim t^{1/3}$, independent of the film-substrate interaction strength $\epsilon_{WA}$, consistent with curvature-driven diffusion governed by the Lifshitz-Slyozov mechanism~\cite{lifshitz,Puri-book,Bray}. Adhesion therefore controls primarily the incubation time for rupture but not the transport mechanism of coarsening.

In the passive film, lateral hole growth [Fig.~\ref{fig:passive-liquid}(b)] follows a power law $\ell_{xy}(t)\sim t^{\alpha}$ with $\alpha\approx0.7$. This behavior is consistent with slip-dominated dewetting, where a constant capillary driving force at the contact line is balanced by frictional dissipation that increases with the size of the dewetted region~\cite{Redon1994,Reiter2000,Khanna2000}. The resulting scaling $\ell_{xy}\sim t^{2/3}$ is well established experimentally and theoretically for thin viscous films, and our measured exponent lies within the reported range ($0.63$-$0.71$). This agreement confirms that the passive simulations reproduce the known thin-film rupture dynamics, providing a controlled reference for identifying activity-induced deviations.

\begin{figure*}
	\centering
	\includegraphics[width=1.0\textwidth]{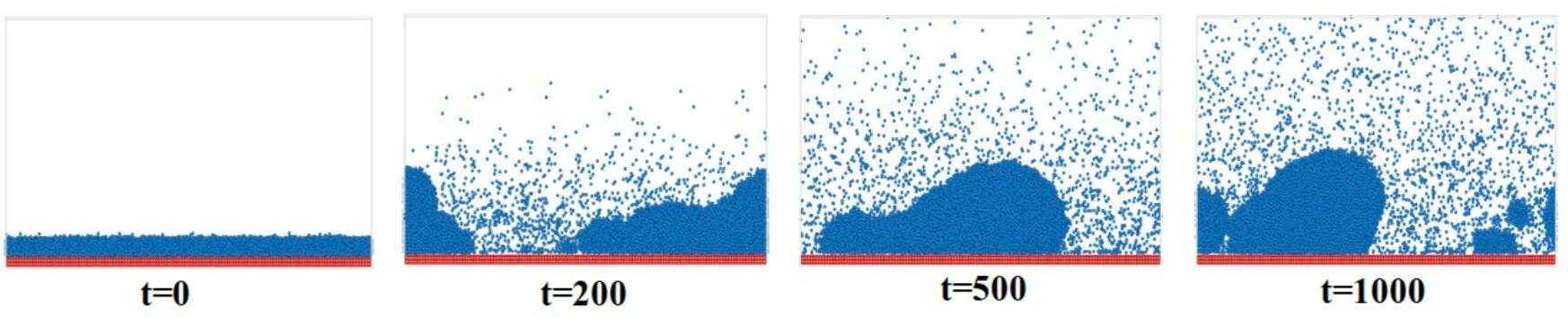}
	\caption{Side view of an active thin film undergoing dewetting for activity $f_A=1$ and film-substrate interaction $\epsilon_{WA}=0.5$. Red particles denote the substrate and blue particles the active fluid. The film develops protrusive structures and partial detachment from the substrate, indicating that active stresses compete with adhesion.}
	\label{fig:active1}
\end{figure*}
\textit{Active films.}-
Introducing activity qualitatively restructures the dewetting instability. 
Representative configurations (Fig.~\ref{fig:active1}) show that the film no longer forms compact domains characteristic of passive rupture. Instead, elongated protrusions and partial lifting of the film from the substrate appear, particularly for weak film-substrate interaction. This morphology indicates that internally generated active stresses locally overcome adhesion-mediated stabilization.

The active domains develop flattened, sheet-like protrusions with enhanced interfacial curvature at their leading edges, reflecting direct reshaping of the interface by persistent active stresses. Unlike passive films, where adhesion determines the incubation time for rupture, increasing activity progressively suppresses this adhesion-controlled delay: at sufficiently high activity, rupture times become nearly independent of substrate interaction strength. Activity therefore not only accelerates dewetting but dynamically lowers the effective stabilization barrier, providing a nonequilibrium route to rupture that is distinct from classical wettability-driven mechanisms. The resulting protrusive morphologies resemble those observed in spreading cellular layers~\cite{Beaune}, suggesting that activity-adhesion competition alone is sufficient to generate biologically relevant interfacial instabilities.
\begin{figure*}
	\centering
	\begin{subfigure}{0.32\textwidth}
		\centering
		\includegraphics[width=\linewidth]{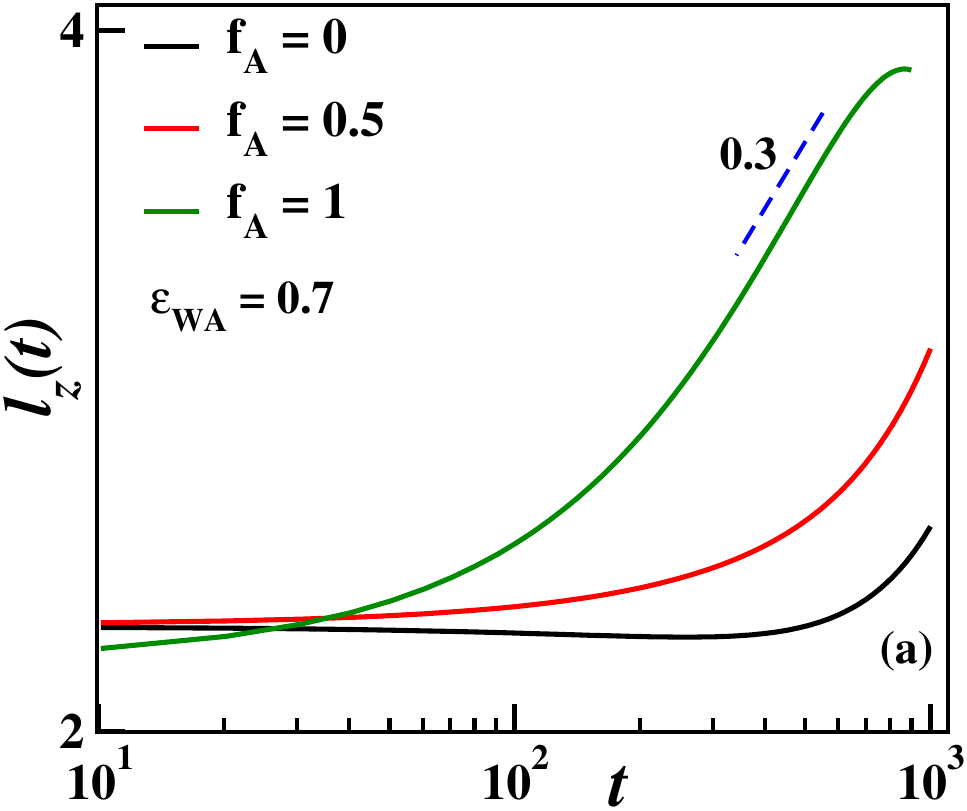}
	\end{subfigure}
	\hfill
	\begin{subfigure}{0.32\textwidth}
		\centering
		\includegraphics[width=\linewidth]{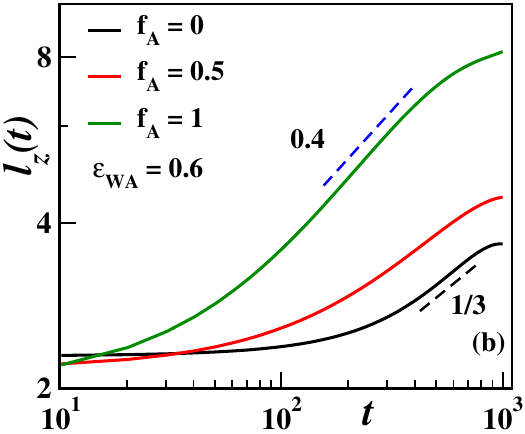}
	\end{subfigure}
	\hfill
	\begin{subfigure}{0.32\textwidth}
		\centering
		\includegraphics[width=\linewidth]{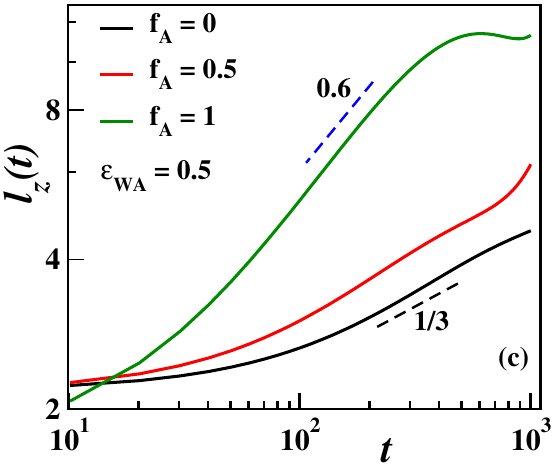}
	\end{subfigure}
	\caption{Vertical liquid accumulation length $\ell_z(t)$ for different activity strengths and film-substrate interactions. The growth exponent increases continuously from diffusion-limited behavior ($\alpha\approx1/3$) at strong adhesion to enhanced growth ($\alpha\approx0.6$) at weak adhesion and high activity. Dashed lines indicate power-law guides to the eye.}
	\label{fig:lengthscale}
\end{figure*}

The primary effect of activity is revealed by the coarsening dynamics. Figure~\ref{fig:lengthscale} shows the evolution of the vertical liquid accumulation length $\ell_z(t)$ for different adhesion strengths and activities. 
For strong adhesion ($\epsilon_{WA}=0.7$), activity advances the onset of rupture but the subsequent growth remains diffusion limited, $\ell_z(t)\sim t^{1/3}$. Thus adhesion fixes the transport mechanism even in the presence of activity.

For weaker adhesion, however, activity alters the growth law itself. At intermediate interaction strength ($\epsilon_{WA}=0.6$) the exponent increases to $\alpha\approx0.4$, while for weak adhesion ($\epsilon_{WA}=0.5$) and high activity ($f_A=1$) the exponent approaches $\alpha\approx0.6$.  The systematic increase from $\alpha\simeq1/3$ to $\alpha\simeq0.6$ indicates a crossover from curvature-controlled diffusive transport to persistence-driven transport produced by active motion.

Activity generates a collective persistence length $\ell_p \sim \hat{v}_{r_c}\tau_p$ associated with alignment-induced coherent motion. When the persistence length is small compared to the domain size ($\ell_p \ll \ell_z$), alignment decorrelates within domains and transport remains effectively diffusive. As activity increases and the persistence length becomes comparable to or larger than the domain size ($\ell_p \gtrsim \ell_z$), coherently moving clusters traverse domains before losing alignment, producing advective mass transport toward interfaces. Growth therefore crosses over from diffusion-limited to persistence-assisted transport, leading to exponents larger than $1/3$ and consistent with the measured $\alpha \approx 0.6$ in the weak-adhesion, high-activity regime.

\begin{figure}
	\centering
	\includegraphics[width=0.95\linewidth]{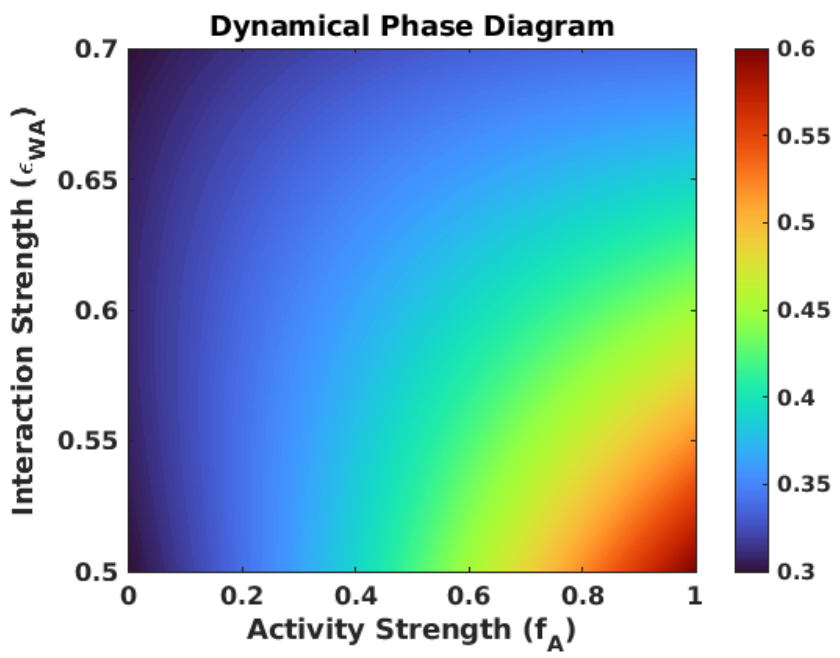}
	\caption{Growth exponent $\alpha$ as a function of activity strength $f_A$ and film-substrate interaction $\epsilon_{WA}$. The crossover from diffusion-dominated to activity-dominated growth is controlled by the competition between persistence and adhesion.}
	\label{fig:phasediagram}
\end{figure}
The dependence of the exponent on activity and adhesion is summarized in Fig.~\ref{fig:phasediagram}. 
The coarsening dynamics are governed by a competition between persistence generated by activity and stabilization by substrate adhesion. Low activity or strong adhesion yields diffusion-dominated growth, whereas increasing activity progressively shifts the system into an activity-dominated regime with faster transport. The smooth variation of $\alpha$ demonstrates a continuous nonequilibrium crossover rather than a sharp transition.

\begin{figure}
	\centering
	\includegraphics[width=0.49\linewidth]{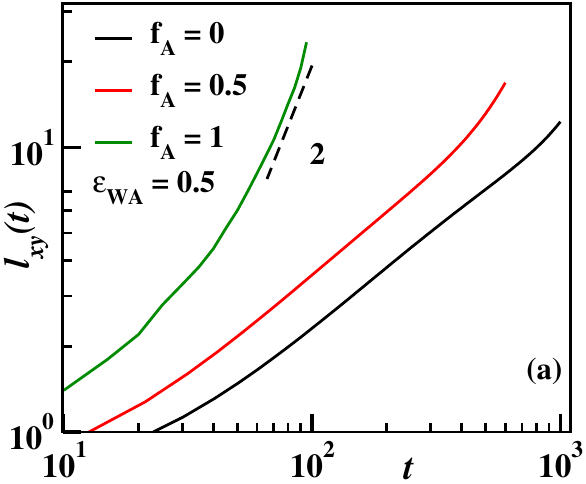}
	\includegraphics[width=0.49\linewidth]{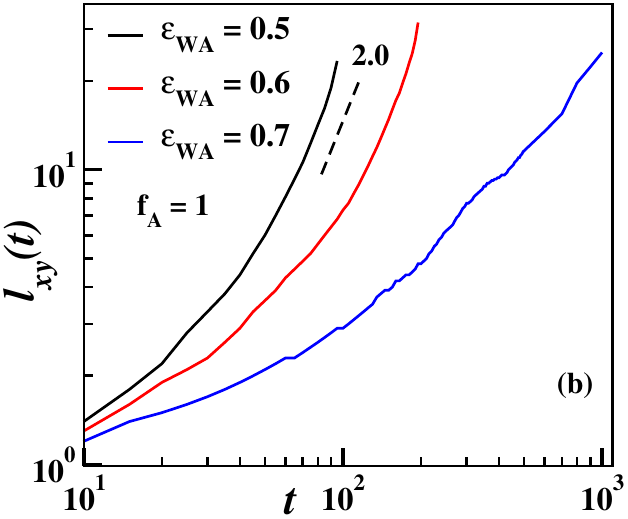}
	\caption{Lateral rupture propagation length $\ell_{xy}(t)$. (a) Dependence on activity at fixed interaction. (b) Dependence on interaction at fixed activity. High activity under weak adhesion produces strongly accelerated spreading of the rupture front.}
	\label{fig:active_rupture}
\end{figure}

Activity also modifies rupture-front propagation. Figure~\ref{fig:active_rupture} shows the evolution of the lateral rupture length $\ell_{xy}(t)$. In passive films, lateral spreading is governed by a capillary driving force at the contact line balanced by frictional dissipation, yielding slip-dominated growth consistent with $\ell_{xy} \sim t^{2/3}$. Under high activity and weak adhesion, persistent active stresses enhance contact-line motion, leading to accelerated super-linear spreading. Increasing adhesion counteracts this active forcing and restores friction-limited propagation.

Taken together, these results reveal a qualitative change in the governing physics. In passive films, bulk coarsening ($\ell_z$) follows curvature-driven diffusive relaxation after spinodal rupture, whereas rupture spreading ($\ell_{xy}$) is governed by capillary-driven contact-line motion along the substrate. In active films they respond differently: activity enhances bulk transport while independently accelerating interfacial propagation. This decoupling demonstrates that activity does not merely renormalize effective surface forces but generates a distinct nonequilibrium interfacial instability controlled by the competition between persistence and adhesion.\\


\textit{Conclusion.}-
We investigated dewetting of a thin film composed of active particles and demonstrated that internally generated persistence fundamentally alters the instability governed by substrate adhesion. In passive films, adhesion primarily determines the rupture time while the post-rupture growth remains diffusion limited. Activity breaks this universality: as persistence increases, the growth exponent of the accumulating liquid domains continuously rises above the Lifshitz-Slyozov value and the rupture front becomes strongly accelerated. 

Importantly, the vertical accumulation length and the lateral rupture propagation respond differently to activity, revealing a decoupling between bulk transport and interfacial spreading that has no passive counterpart. The resulting dynamics are controlled by a competition between active persistence and adhesion, rather than by surface forces alone. 

These findings show that activity does not merely renormalize effective interactions but generates a distinct nonequilibrium interfacial instability. Active thin films therefore constitute a minimal model for internally driven rupture processes, relevant to dewetting of active suspensions, biofilms, and cellular layers.\\

\textit{Acknowledgements.}~~B.S.G. acknowledges the Science and Engineering Research Board (SERB), Department of Science and Technology (DST), Government of India (No. CRG/2022/009343) for financial support. P.M. acknowledges the support by the INSPIRE Fellowship (No. DST/INSPIRE Fellowship/2023/IF230398) of the Department of Science and Technology (DST), Government of India. D.D. acknowledges CSIR, India [09/0844(23194)/2025-EMR-I] for SRF fellowship.

\clearpage
\onecolumngrid

\begin{center}
	{\Large \textbf{Supplemental Material}}

\end{center}

\setcounter{equation}{0}
\renewcommand{\theequation}{S\arabic{equation}}
\setcounter{figure}{0}
\renewcommand{\thefigure}{S\arabic{figure}}
\setcounter{NAT@ctr}{0}
\renewcommand{\bibnumfmt}[1]{[S#1]}
\renewcommand{\citenumfont}[1]{S#1}

\subsection{Active Dynamics and Simulation Details}

Self-propulsion is implemented through a Vicsek-type alignment rule. Each mobile particle $i$ experiences an active force

\begin{equation}
	\mathbf{F}^{S}_{i}=f_A \hat{\mathbf{v}}_{r_c},
\end{equation}

where $f_A$ is the activity strength and $\hat{\mathbf{v}}_{r_c}$ is the normalized average velocity of neighboring particles within radius $r_c$,

\begin{equation}
	\hat{\mathbf{v}}_{r_c}=
	\frac{\sum_{j} \mathbf{v}_j}{\left|\sum_{j} \mathbf{v}_j\right|}.
\end{equation}

To avoid artificial heating, activity modifies only the direction of motion. If $\mathbf{v}_i^{\rm pas}(t+\Delta t)$ is the velocity obtained from the passive Langevin update, the active velocity is assigned as

\begin{equation}
	\mathbf{v}_i(t+\Delta t)=
	|\mathbf{v}_i^{\rm pas}(t+\Delta t)|
	\frac{\mathbf{v}_i^{\rm pas}(t+\Delta t)+\frac{\mathbf{F}_i^S}{m}\Delta t}
	{\left|\mathbf{v}_i^{\rm pas}(t+\Delta t)+\frac{\mathbf{F}_i^S}{m}\Delta t\right|}.
\end{equation}
This procedure preserves the instantaneous kinetic energy while introducing persistence in particle motion.

The system consists of a three-dimensional box of size $L_x=L_y=128$ and $L_z=70$ (in units of particle diameter $\sigma$), with periodic boundary conditions in the lateral directions and a reflecting boundary normal to the substrate. The solid substrate is modeled as a slab of immobile particles of thickness $3\sigma$ spanning the entire $x$-$y$ plane. Fluid particles are initially placed randomly above the substrate to form a uniform film of thickness $3\sigma$ at an average density $\rho=0.8$.
\begin{figure}[h]
	\centering
	\includegraphics[width=0.5\linewidth]{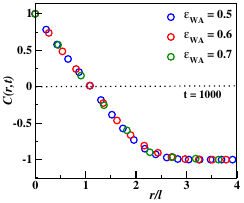}
	\caption{Scaled correlation functions for different substrate interactions. Spatial separation is rescaled by the instantaneous domain size.}
	\label{fig:scaledcorr}
\end{figure}

Molecular dynamics simulations are employed to integrate the equation of motions using the velocity-Verlet algorithm with time step $\Delta t=0.001$.  For convenience we use $m=\sigma=\epsilon=k_B=1$, $k_B$ being the Boltzmann constant. Temperature is fixed at $T=0.6$ by the Langevin thermostat.  Length, temperature and time are measured in units of $\sigma$, $\epsilon/k_B$ and $(m\sigma^2/\epsilon)^{1/2}$ respectively. All results are averaged over $30$ independent initial conditions.

\subsection{Order Parameter and Domain Size Extraction}

A coarse-grained density order parameter is defined as

\begin{equation}
	\psi(\mathbf{r},t)=
	\begin{cases}
		0, & \text{substrate} \\
		+1, & \rho(\mathbf{r},t)>\rho_c \\
		-1, & \text{rupture region}
	\end{cases}
\end{equation}
The characteristic length scale $\ell(t)$ is obtained from the first zero crossing of $C_{\psi\psi}(r,t)$.

When distances are rescaled by $\ell(t)$, correlation functions collapse onto a master curve (Fig.~\ref{fig:scaledcorr}), confirming self-similar coarsening and validating the single-length-scale description.

\subsection{Morphology Snapshots}

\begin{figure*}[t]
	\centering
	\includegraphics[width=0.75\textwidth]{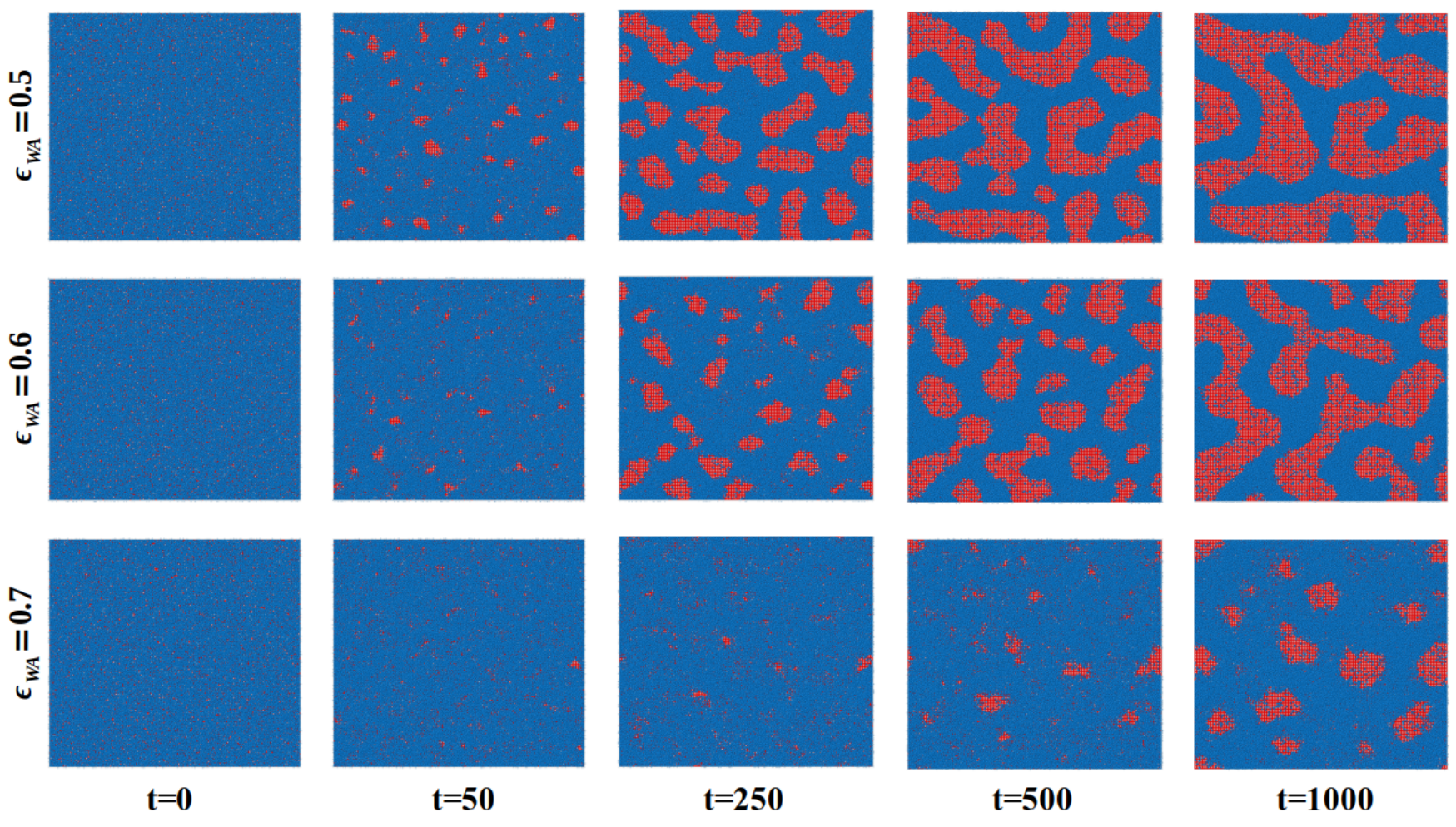}
	\caption{Top view of passive ($f_A=0$) thin films during dewetting for the film-substrate interaction $\epsilon_{WA}=0.5$, illustrating rupture and lateral hole growth. Red and blue particles denote the substrate and fluid respectively.}
	\label{fig:Passive1}
\end{figure*}

Figure~\ref{fig:Passive1} shows representative passive configurations. Increasing film-substrate interaction delays rupture but does not alter the qualitative morphology.

\begin{figure*}[t]
	\centering
	\includegraphics[width=0.75\textwidth]{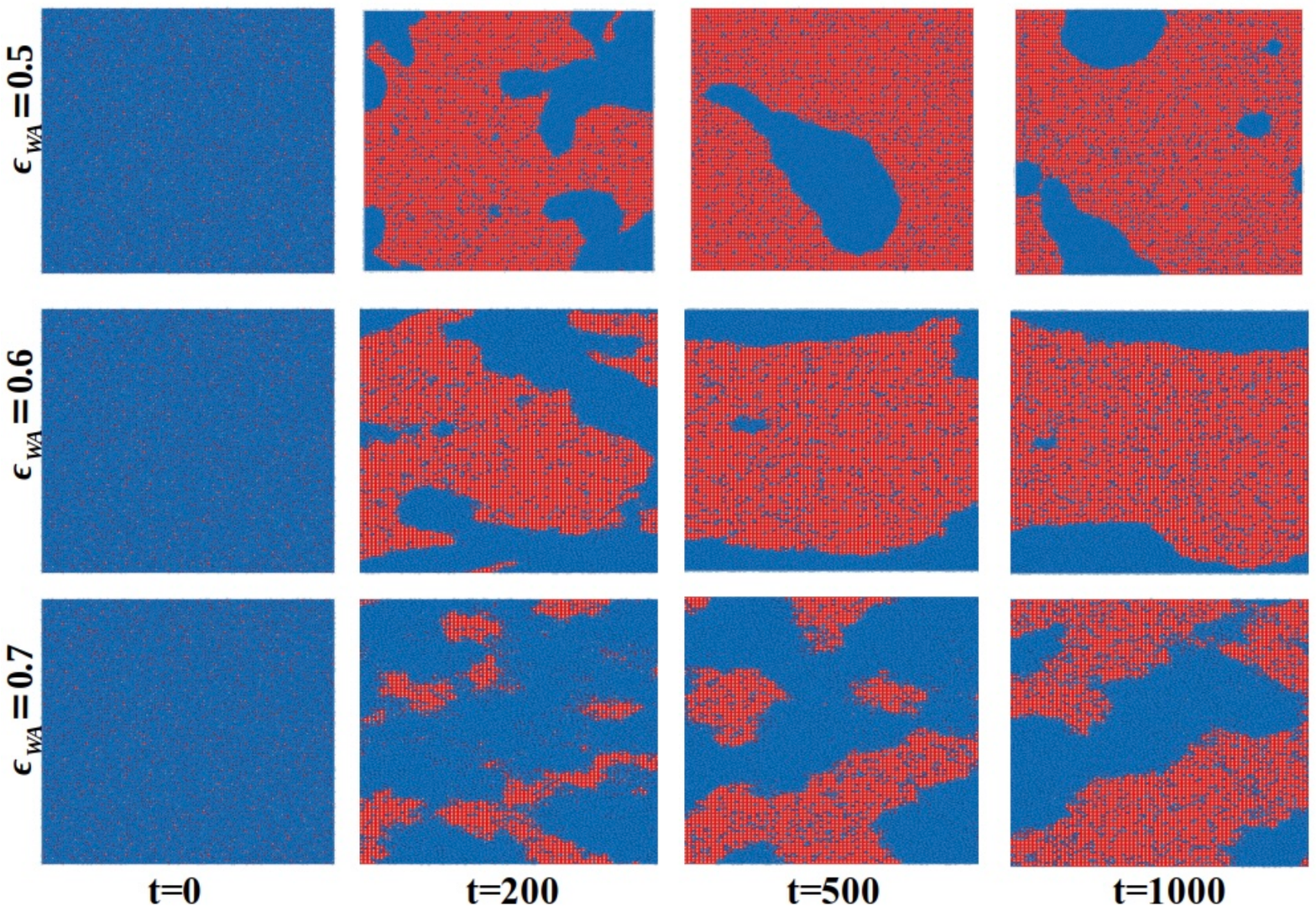}
	\caption{Top-view configurations of active films ($f_A=1$) for varying film-substrate interaction strengths $\epsilon_{WA}$, illustrating activity-induced morphological changes during rupture. Red and blue particles represent the substrate and fluid, respectively.}
	\label{fig:active2}
\end{figure*}

Figure~\ref{fig:active2} shows the corresponding active systems.  Activity reduces the rupture time and promotes partial detachment from the substrate.

\subsection{Density Profiles}

\begin{figure}[h]
	\centering
	\includegraphics[width=0.4\linewidth]{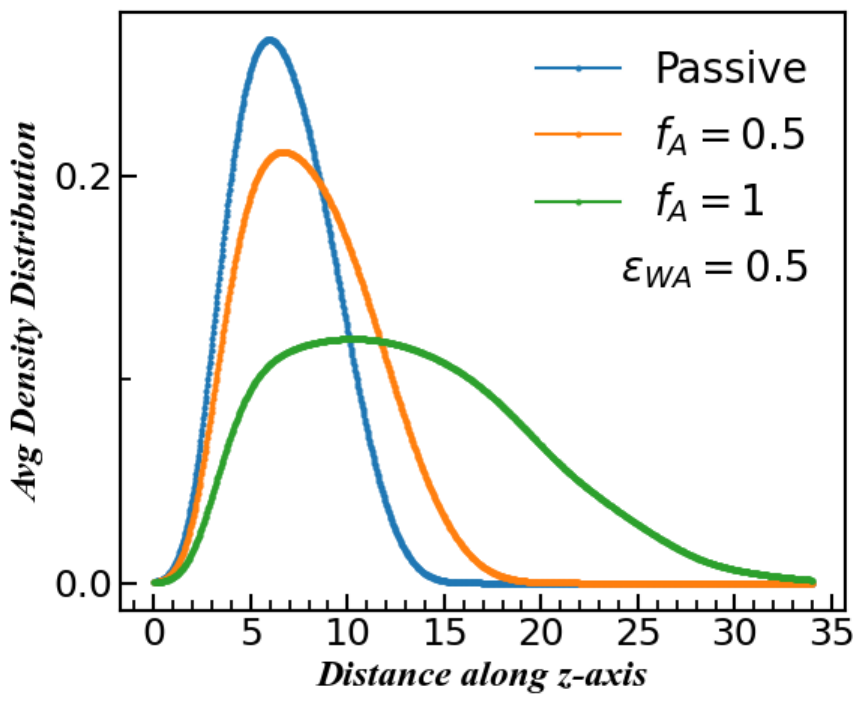}
	\includegraphics[width=0.4\linewidth]{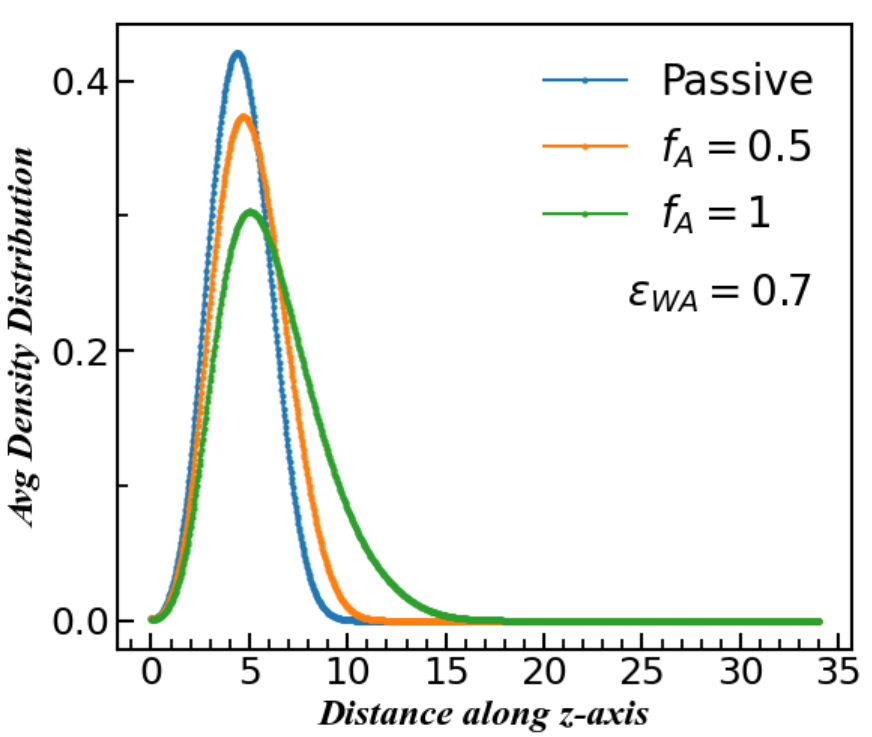}
	\caption{Time-averaged density profiles normal to the substrate for weak and strong adhesion.}
	\label{fig:density}
\end{figure}

The density profiles (Fig.~\ref{fig:density}) broaden with increasing activity, indicating redistribution of particles away from the substrate. The effect is stronger for weak film-substrate interaction.

\subsection{Contact Angle Measurement}

Assuming a spherical-cap droplet, the apparent contact angle is calculated from droplet height $h$ and base radius $r$,

\begin{equation}
	\theta = 2 \tan^{-1}\left(\frac{h}{r}\right).
\end{equation}

\begin{figure*}[t]
	\centering
	\includegraphics[width=0.3\textwidth]{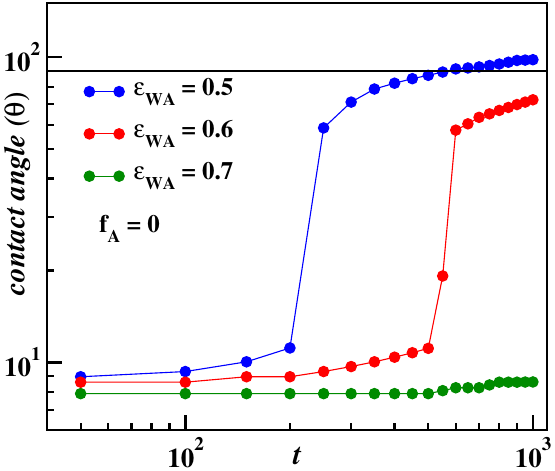}
	\includegraphics[width=0.3\textwidth]{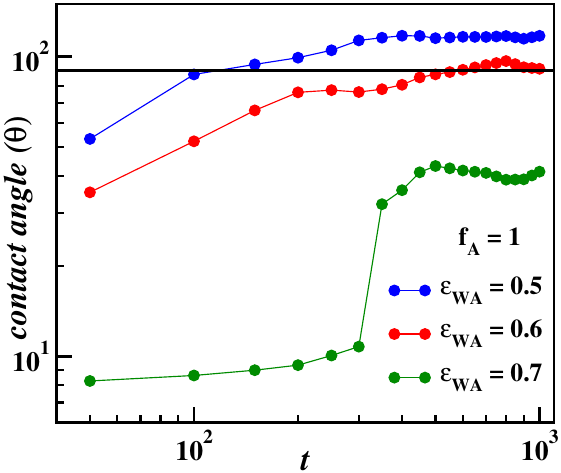}
	\includegraphics[width=0.3\textwidth]{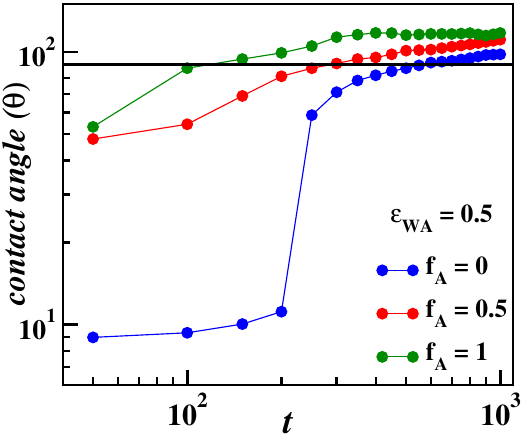}
	\caption{Time evolution of the apparent contact angle (hydrophilic-to-hydrophobic crossover) for (a) passive films, (b) active films at fixed activity, and (c) varying activity at fixed interaction strength.}
	\label{fig:contactangle}
\end{figure*}

Activity accelerates the increase of contact angle and reduces its sensitivity to substrate interaction strength (Fig. \ref{fig:contactangle}).

\end{document}